\documentclass [12pt,twoside] {article}
\newlength{\awidth}
\newlength{\aheight}
\newlength{\uswidth}
\newlength{\usheight}
\newlength{\spacing}
\newlength{\topoff}
\newlength{\margoff}
\newlength{\margin}
\newlength{\hmargin}
\font\vbig=cmbx10 scaled\magstep3
\font\big=cmbx10 scaled\magstep1

\newcommand{\citen}[1]{
  \cite{#1}}

\newenvironment{eqtn}[1]{
  \begin{equation}
    \label{#1}
}{
  \end{equation}}
\newlength{\subfigwidth}
\newcounter{fignr}
\newcounter{subfignr}[fignr]
\renewcommand{\thesubfignr}{\alph{subfignr}}

\begin{document}
\setlength{\baselineskip}{\spacing}

\begin{titlepage}
\begin{flushright}
USITP-95-7\\
Aug, 1995\\
\end{flushright}
 \parindent 0pt
\null
\vskip 2cm
\begin{center}
 {\vbig On Self--Similar Global Textures in an expanding Universe}
 \vskip 1.5cm
{\bf Stefan {\AA}minneborg\footnote{E-mail: {\em stefan@vana.physto.se}} and
Lars Bergstr\"om}\footnote{E-mail: {\em lbe@vana.physto.se}}\\
 \vskip 1.5cm
 Department of Physics\\
 University of Stockholm\\
 Vanadisv\"agen 9\\
 S--113 46 Stockholm, Sweden\\
 \end{center}
 \vskip 1.5cm
 \begin{abstract}
We discuss self-similar solutions to $O(4)$ textures in Minkowski space and
 in flat Friedmann-Robertson-Walker backgrounds.
We show that  in the Minkowski case
there exist no solutions with winding number greater than unity.
However, we find besides the known solution with unit winding number also
 previously unknown solutions corresponding to
winding number less than one.
The applicability of the non-linear sigma model approximation
 is discussed.
We point out that no spherically symmetric exactly self-similar solutions exist
 for radiation or matter dominated FRW cosmologies, but we
 find a way to relax the assumptions of self-similarity that give
 us approximative solutions valid on intermediate scales.
\end{abstract}
\end{titlepage}
\newpage
\font\big=cmbx10 scaled\magstep1
\newcommand{\beq}{\begin{equation}}
\newcommand{\eeq}{\end{equation}}
\newcommand{\beqa}{\begin{eqnarray}}
\newcommand{\eeqa}{\end{eqnarray}}
\newcommand{\eps}{\varepsilon}
\newcommand{\del}{\partial}
\newcommand{\figu}{Fig.\,~}
\newcommand{\tab}{Tab.\,~}
\newcommand{\emline}{\vspace{.4cm}}

\section{Introduction}

Since some time, it has been realized \cite{Turok} that defects (textures)
 associated with
the non-trivial winding of massless scalar fields may be of interest even
though they are intrinsically unstable if the winding number is large enough
\cite{Ryden,Borrill,Stefan}. Indeed, it is the fact that textures
continually enter the horizon during the evolution of the universe that
makes the spectrum of density fluctuations near scale-invariant (although
not Gaussian) and
makes textures promising candidates for large scale structure formation
even in light of the COBE results \cite{COBE,Turok_Stock}.

A simple theory that admits global textures is given by the lagrangian
\begin{eqtn}{lagr}
 {\cal L}=
	\frac{1}{2} \partial_\mu {\bf \Phi} \cdot \partial^\mu {\bf \Phi}
	- \frac{\lambda}{4}( {\bf \Phi}^2 - \phi_0^2 )^2
\end{eqtn}
 where
 ${\bf \Phi}$
 is a $4$ component real scalar field,
Here $\phi_0$ is the symmetry breaking scale
 and $\lambda$ is a dimensionless coupling constant.
For convenience we do the rescaling ${\bf \Phi} \rightarrow \phi_0 {\bf \Phi}$,
 so the action for this theory will be
\begin{eqtn}{action}
 S[{\bf \Phi}] = \phi_0^2 \int d^4x \sqrt{-g}
	\left( \mbox{$1\over2$}
	 \partial_\mu {\bf \Phi}  \cdot   \partial^\mu {\bf \Phi}
	 - \mbox{$w\over4$}( {\bf \Phi}^2 - 1 )^2  \right),
\end{eqtn}
 were $w=\lambda \phi_0^2$ and  $g$ is the determinant of the space-time
metric.

Upon quantization we have in this theory one massive Higgs particle with mass
 $m_H=\sqrt{w/2}$ and $N-1$ massless (Goldstone) bosons.
We will however not be interested in the particle spectrum, instead we will
 only consider the classical equations of motion, for the field ${\bf \Phi}$,
\begin{eqtn}{eom1}
 \partial_\mu( \sqrt{-g} \partial^\mu {\bf \Phi} )
	= -  \sqrt{-g} w ( {\bf \Phi}^2 - 1 ) {\bf \Phi}.
\end{eqtn}
For cosmological applications, it is important that the Goldstone modes
remain massless, creating long-range correlations and field dynamics
over cosmologically relevant length scales.
Arguments, based on
quantum gravity effects, have been given \cite{Kamion}
which seem to make the survival of exact global symmetries questionable.
This statement is, however,  based on unknown physics at the Planck scale.
More specifically, mechanisms present, e.g. in string models may
well protect the potential of the Goldstone modes of the
texture scalar fields.
(For a recent discussion of such mechanisms, see \cite{Kallosh}.)

Here we will not enter into this discussion but simply
assume that textures can exist and study the properties of their
 evolution when the dynamics is given by the action (\ref{action}).

We will parameterize the field using hyper-spherical coordinates
 $\rho$, $\chi$, $\tilde{\theta}$
 and $\tilde{\varphi}$, in the following way,
\begin{eqtn}{field}
 { \bf \phi} ({\bf r},t) = \rho ( \cos \chi
 , \sin \chi \cos \tilde{\theta}
 , \sin \chi  \sin \tilde{\theta} \cos \tilde{\varphi}
 , \sin \chi  \sin \tilde{\theta} \sin \tilde{\varphi} ).
\end{eqtn}
We will look at the ``spherically symmetric'' (or hedgehog) ansatz,
were we let the
 coordinate functions depend on time $t$ and the spatial
 spherical coordinates $r$, $\theta$ and $\varphi$
 as $\rho=\rho(r,t)$, $\chi=\chi(r,t)$,
 $\tilde{\theta}=\theta$ and $\tilde{\varphi}=\varphi$.

It is common to consider the field as a stiff source, which means that
 one is assuming that the self-coupling of the field is much stronger
 than the self-gravitational coupling.
Thus only the background metric is required in the equation of motion
 for the field.
The perturbation of the background metric can then be calculated from
 Einstein's equations with the stress-energy tensor of the texture field
 added.
The applicability of the stiff approximation  in the self-similar case
 is discussed in \cite{selfgrav}.

When one studies the formation of large scale structure in the early universe,
 the background metric is taken to be Friedmann-Robertson-Walker (FRW).
We will discuss how one can find solutions valid at medium large scales
  using the equation for  the Minkowski background.

\section{Minkowski background}

For a Minkowski background the equations of motion~(\ref{eom1}) in terms of the
hyper-spherical coordinates become
\begin{eqtn}{eomM1}
 \begin{array}{l}
  	r^2(\rho^2 \dot{\chi})\dot{ } = (\rho^2 r^2  \chi')'
	-  \rho^2 \sin 2\chi \\
        r^2 w ( \rho^2 - 1 ) =
        r^2( \dot{\chi}^2 - \chi'^2) - 2 \sin^2 \chi
        + (( r^2  \rho')' - r^2 \ddot{\rho})/\rho.
 \end{array}
\end{eqtn}
For length scales larger than the inverse mass of the radial
``Higgs'' mode $m_r^{-1}=(\lambda \phi_0^2)^{-1/2}$
the dynamics of the field can be described by a nonlinear $\sigma$
model (NLSM) \cite{Turok}.
The NLSM is characterized by that the field is exactly on
 the vacuum-manifold everywhere, thus $\rho=1$.
In this case the first equation of (\ref{eomM1}) admits a
 self-similar ansatz $y=r/t$ and becomes
\begin{eqtn}{semiy}
 (1-y^2)(y^2\chi_{yy}+2y\chi_y)=\sin 2\chi(y).
\end{eqtn}
This equation has a singular behavior at $y=0$ and $y=\pm 1$,
 the conditions for regular solutions are
 $\sin 2\chi(0)=0$ and  $\sin 2\chi(\pm 1)=0$.

This equation  has the well known solutions found by Turok and
Spergel \citen{Spergel}.
\begin{eqtn}{atan}
 \chi(y) = m \pi \pm 2 \:\arctan (\pm y),
\end{eqtn}
 where $m$ is an integer.

These solutions are indeed very special, coming from
the spherically symmetric self-similar ansatz to the non-linear sigma model
approximation.
An important feature is, however, that these  symmetric scaling
solutions appears
in the numerical simulations as attractors \cite{Ryden}. In fact, many of
the calculations of the effects of textures, e.g. on the microwave background
radiation, rely on the use of these simple analytic solutions
\cite{Spergel,Durrer}.

In order to see when the NLSM approximation is applicable
 we go a step beyond it.
Instead of insisting on $\rho\equiv 1$, we will assume only
 that the derivatives of $\rho$ are negligible in the equations of motion.
We will then recover the NLSM equation for $\chi$, so for $\chi$ we will
 use  the selfsimilar NLSM solution~(\ref{atan}).
The second  of the eqs. (\ref{eomM1}) $\rho$ becomes
\begin{eqtn}{rho1}
  \rho^2 = 1 + ( \dot{\chi}^2 - \chi'^2 - \frac{2}{r^2} \sin^2 \chi )/w,
\end{eqtn}
 which gives
\begin{eqtn}{rho2}
\rho^2 = 1 - \frac{12 t^2 - 4 r^2}{w(r^2 + t^2)^2},
\end{eqtn}
 upon insertion of our self-similar NLSM solution for $\chi$.
It can be checked that $\rho'$ and $\dot{\rho}$ can be neglected in the first
  of the eqs. (\ref{eomM1}) if $r^2 + t^2 >> 1/w$.
Thus we conclude that there exists a $r_0>>1/\sqrt{w}$ such that
 the selfsimilar solution~(\ref{atan}) is valid
 for all $r^2 + t^2 > r_0$.

If we have a field that initially for $t<0$, $t^2>>1/w$ is described by
 $ \chi(r,t) = 2 \:\arctan (-r/t)$ we will have an unwinding event for
 $r^2 + t^2 < r_0^2$
 where the field is forced to leave the vacuum manifold.
The solution~(\ref{atan}) is valid right to the time $t=r-r_0$,
 when the information from the unwinding event reaches $r$.
We thus can match  the solution at $t=0$, $\chi(r,0)=\pi$,
 $\dot{\chi}(r,0) = 2/r$,
 with  $\chi(r,t) = 2 \pi - 2 \:\arctan ( r/t)$, valid for $0<t<r-r_0$.

How the field behaves for $t>r-r_0$ depends on the details of the
 unwinding event and must be decided by making
 a numerical simulation of the full field equations \cite{Barriola}.
One thus find that the field after the unwinding  goes
 asymptotically to the NLSM solution
 $ \chi(r,t) = \pi + 2 \:\arctan ( r/t)$ for $t>r$.
This solution describes an expanding shell of goldstone bosons,
 the winding number is zero.

Before we continue discussing Minkowskian  self-similar solutions,
 we study the equations for the FRW background metric.

\section{FRW background}

We will  discuss the evolution of spherical textures
 in a  flat FRW background
\beq
ds^2=a^2(\eta)(d\eta^2-dr^2-r^2(d\theta^2+sin^2\theta d\varphi^2)).
\eeq
Here $\eta$ is the conformal time.
The time dependence for $a$
is $a(\eta)\propto \eta^\alpha$, where $\alpha=1$ corresponds to a
radiation dominated universe and $\alpha=2$ corresponds to a
 matter dominated universe.

The equation of motion for the NLSM is now
\beq\label{whole}
 (r^2\chi_{r})_r - r^2( \chi_{\eta \eta}+2\frac{\alpha}{\eta}\chi_\eta )
 = \sin 2\chi
\eeq
We are interested in solutions where unwinding can occur so we
 try  the selfsimilar ansatz $\chi(r,t)=\chi(r/t)$, where $t=\eta-\eta_*$
 and  $\eta_*$ is the time for the unwinding.
The equations with $\alpha\neq0$  admit this ansatz only if
 $\eta_*=0$ which coincides with the time for the big bang singularity.
One would nevertheless hope to have some use of this ansatz if we are
 interested in expanding textures that  unwinded very early.

The self-similar ansatz $y = r/\eta$ gives
\beq\label{sege}
y^2(1-y^2)\chi_{yy}+2y(1+y^2(\alpha-1))\chi_y=\sin 2\chi(y).
\eeq
However, we will show that with $\alpha = 1\ {\rm or}\ 2$
 there does not exist any  non-trivial
solutions to (\ref{sege}) passing through $y=1$ with finite derivative.

To show this we use the regularity conditions  that we get by
 letting $y \rightarrow 1$ in
 (\ref{sege}) and  the first and second derivative of that equation.
We thus have at $y=1$
\begin{eqtn}{regRW}
 \begin{array}{l}
  2 \alpha \chi_y = \sin 2 \chi, \\
  (\alpha - 1) \chi_{yy} + (3 \alpha - 2 - \cos 2 \chi) \chi_y = 0, \\
  (\alpha - 2) \chi_{yyy} + (6 \alpha - 9 - \cos 2 \chi) \chi_{yy}
   + (6 (\alpha-1) +2 \cos 2 \chi) \chi_y = 0. \\
 \end{array}
\end{eqtn}
\emline

For $\alpha = 1$ we find $(\cos 2\chi(1) - 1)\chi_y(1) = 0$.
This gives $\chi_y(1)=0$ since  $\cos 2\chi(1)=1$ implies
 $\chi_y(1)=0$.
For $\alpha = 2$ we find
$$
(\cos^2 2\chi(1) - {14\over 3}\cos 2\chi(1) + {11\over 3})\chi_y(1) = 0,
$$
 the two solutions for $\cos 2 \chi(1)$ are
$\cos 2 \chi(1) = {7\pm 4 \over 3}$ so the only real solution is also here
 $\chi_y(1)=0$.

We see that in both cases for regular solutions we must have
 $\chi_y(1)=0$.
Since $\chi(y) = n\pi/2$, integer $n$  are solutions to (\ref{sege})
 with $\chi_y(1) = 0$ and $\sin 2\chi(1)=0$ we conclude that there does
 not exist any non-trivial regular solutions.

Instead we go over to discuss the validity of using the equation for
 the Minkowski background as the limiting case when we look at small
 scales.
If we substitute $\eta = \eta_* + t$ in (\ref{whole})
 and assume that $t<<\eta_*$ we get
\beq\label{small}
 (r^2\chi_{r})_r - r^2( \chi_{tt}+2\frac{\alpha}{\eta_*}\chi_t )
 = \sin 2\chi.
\eeq
Now if we can neglect the term linear in $\chi_t$ compared to $\chi_{tt}$
 we recover the Minkowski equation.
Inserting the solution $\chi(r/t)=2 \arctan (r/t)$ we  see
 that this approximation is consistent only when
 $|t| \eta_* >> \alpha (r^2+t^2)$.
Thus  we have to try a different approximation in order to get something
 valid for $t\approx 0$.

We have found a way to get rid of the term linear in the time-derivative
 by a  certain transformation.
In the case $\alpha=1$ when the equation of motion is
\beq
 (r^2\chi_{r})_r - r^2( \chi_{\eta \eta}+\frac{2}{\eta}\chi_\eta )
 = \sin 2\chi
\eeq
we can make the substitution
 $\chi(r,\eta) = {\eta_* \over \eta}\psi(r,\eta)$ and get
\beq\label{rad1}
 (r^2\psi_{r})_r - r^2 \psi_{\eta \eta}
 = {\eta \over \eta_*} \sin 2{\eta_* \over \eta}\psi,
\eeq
 which become the Minkowskian equation after substituting
 $\eta=\eta_* + t$ and neglecting $t$ compared with $\eta_*$.
Thus we find for $\alpha=1$ the solutions
\begin{eqtn}{solrm}
\chi(r,\eta) = {\eta_* \over \eta_*+t} \psi_M(r/t)
\end{eqtn}
 valid for $t<<\eta_*$,
 where $\psi_M(r/t)$ is any  solution to the Minkowskian equation.

The same trick can be done in the case $\alpha=2$,
 but first we have to change to the coordinates
 $u=3\eta_*^2 r$ and $\tau=\eta^3$ in eq. (\ref{whole}) giving
\beq\label{mat1}
 (u^2\chi_{u})_u - (\tau/\tau_*)^{4/3}
 u^2( \chi_{\tau \tau} + \frac{2}{\tau}\chi_\tau )
 = \sin 2\chi,
\eeq
 which has the solutions
\begin{eqtn}{solmm}
\chi(u,\tau) = {\tau_* \over \tau_*+t} \psi_M(u/t)
\end{eqtn}
valid for $t<<\tau_*$.

In the linear approximation of Einstein's eqs.
 one can calculate the Newtonian gravitational acceleration from the
 unwinding texture solution (\ref{atan}), the result is
\begin{eqtn}{force0}
\vec{g}= -  \varepsilon {r \over r^2+t^2}\hat{r},
\end{eqtn}
 where $\varepsilon=8\pi G\phi_0^2$, G the gravitational constant and
 $\hat{r}$ is the radial unit vector.
This acceleration give rise to a velocity kick inwards
 of surrounding homogenous dust of the amount $\pi\varepsilon$
 \cite{Spergel}.

For the solution (\ref{solmm}) the same calculation gives to first order
 in $t/\tau_*$;
\begin{eqtn}{force1}
\vec{g}= - \varepsilon {r \over r^2+t^2}(1-t/\tau_*)\hat{r}.
\end{eqtn}
We notice that the acceleration (\ref{force1}) is enhanced at $t<0$
 compared with the ordinary (\ref{force0}), and vice versa for $t>0$,
 which can be of importance for the form of the resulting matter perturbations.
The velocity kick of the dust over the time interval $\{-t_0,t_0\}$,
 $t_0<<\tau_*$ is $2\varepsilon\arctan t_0/r$, so
 at $r<<t_0$ it is still $\pi\varepsilon$.

\section{New solutions}

We now examine the self-similar solutions of the Minkowskian equation
 of motion in greater detail.

The solutions (\ref{atan}) have a winding charge $Q=\pm1$.
One would perhaps believe that there exist solutions with higher $|Q|$
 than $1$ as has been claimed in \cite{Borrill}, but this is not the case.

For a self-similar spherical texture we have $|Q|\leq 1$.
This is implied by the theorem
 we prove in the appendix
that if $\chi(y)$ is a regular solution to (\ref{semiy}) with $\chi(0)=0$
then $0<|\chi(y)|<\pi$,
for all finite $y$, and we have $0<|\chi(\infty)|\leq\pi$.

The argumentation of reference \cite{Borrill}
 concerning solutions with $|Q|>1$
 is based on the erroneous
 assumption that there exist solutions satisfying the
 boundary conditions $\chi(0)=0$, $\chi(1)=n\pi/2$, with $n>1$,
 (see the corollary of Lemma~1).

We also want to emphasize that
 a boundary value problem such as (\ref{semiy}) with $\chi(0)=0$ and
$\chi(1)=\pi/2$, does not necessarily  possess a unique solution.

Actually we have by numerical means been able to demonstrate
 the existence of whatseems to be  a countably infinite set of
 additional solutions with total winding charge $Q$ less than unity.
These solutions are characterized by
 the number of oscillations around the value $\chi =\pi/2$, and have rapidly
 increasing derivatives at the origin.

We want to demonstrate the existence of these
 new solutions by accurate numerical techniques
 \cite{num} with some modifications necessary
 to handle the singular points of the equation.
We want to solve the boundary value problem using the shooting technique.
The most straightforward strategy would then be to consider the initial
 value problem
 $\chi(0)=0,\; \chi_y(0)=\beta$,
 and numerically integrate
 this to the point $y=1$, we denote the solutions by $\chi(y,\beta)$.
The equation $\chi(1,\beta)=\pi/2$ may  now be solved for $\beta$ by trial.
However, because of the singularities it
 becomes numerically impossible
 to start the integration from $y=0$, so we have to modify our method.
We must start the integration at a small distance away from $y=0$ and
 use a series expansion in order to get an appropriate initial
 condition.

By making a series expansion of $\chi(y)$ close to the origin of the form:
\beq
\chi(y)=\sum_{k=0}^{\infty}a_{2k+1}y^{2k+1}
\eeq
 and inserting this into the differential equation Eq. (\ref{semiy}), one can
 find the coefficients $a_3$, $a_5$,... in terms of $a_1=\beta$.
One finds, e.g.,
$$
a_3={2\beta-4\beta^3/3\over 10}
$$
and
$$
a_5={3\beta-3\beta^3+\beta^5\over 35}.
$$
Let us now consider the initial value problem,
\beq\label{init}
\chi(\eps)=\beta \eps+a_3\eps^3+a_5\eps^5,\; \chi_y(\eps)=\beta,
\eeq
and integrate only up to $y=1-\eps$.
For each $\eps$ we may find a $\beta$ such
that
\beq
\chi(1-\eps,\beta)+\eps\chi_y(1-\eps,\beta) = \pi/2,
\eeq
we then
must check that the value of $\beta$ converges when we choose $\eps$ smaller
and smaller. It is also possible to use
 a more accurate extrapolation formula near y=1, one
may again use the form of the original differential equation to write
$$
\chi(1-z)={\pi\over 2}-\gamma z-{\gamma\over2}z^2-{\gamma\over 6}z^3+
{\gamma (1-\gamma^2)\over 18} z^4+...
$$
where $\gamma = \chi_y(1)$. This expression can also be used to continue
the solution past the singular point
y=1.

With this careful treatment of the singular points
of the differential equation, its solution is otherwise straightforward.
For the numerical solution we used a Runge-Kutta method with
adaptive size control.
We have employed this technique and thus discovered a set of such different
$\beta$'s.
In \figu 1 the first four of the solutions corresponding to these $\beta$'s
are displayed, we number them with the mode number $n$ starting
 with $n=0$ for the analytical solution.

These solutions appear to be  very
robust according to various  stability checks we have made
 of our numerical algorithm, so we are confident in the belief
 that the presence of the solutions is not a numerical artefact.
We have also checked that when we vary $\alpha$ in $(\ref{sege})$
around $0$ we still find solutions which approach our solutions in a
continuous way when $\alpha\rightarrow 0$.
 From the conspicuously regular pattern of
the first solutions shown in \figu 1, we conjecture that the number of
solutions is countably infinite.

That self-similar solutions with winding number less
than unity exist is potentially of great importance, since as shown in
numerical simulations \cite{Ryden} and backed by analytical arguments
\cite{Stefan}, configurations with $Q > 1/2$ collapse and contribute to
structure formation. We expect that well inside the horizon where
spacetime is approximately Minkowski our new scaling solutions could play a
dynamical role in structure formation.
These solutions need a very high resolution numerical code to appear in the
 simulations since the derivatives at $y=0$ are very high.
We plan to investigate these questions as well as the attractor
nature of the solutions in future work.
The applicability of the NLSM for the solutions with winding number
 less than one can be examined in the same way as for the analytical solution.
The NLSM is valid for $r^2+t^2>r_0^2$ and we find
 that we get a factor of around hundred extra in $r_0$
 for each mode,  the condition reads $r_0>>100^n/\sqrt{w}$, where
 $n$ is the mode number.
We can use the solutions for
 times $t<r-r_0$  as for the analytical solution, but
 they can not be matched at $t=r$ with any selfsimilar solution,
 thus for $t>r$ the selfsimilarity will necessarily be lost.

To conclude, we have investigated in quite some detail the nature and validity
of the self-similar ansatz to the texture equations of motion.
We have analyzed possible modifications when one goes  beyond the non-linear
sigma model approximation, the Minkowski background approximation, and
the "ground state" arctangent solution.
In future work, the effects caused
by including the self-gravitational coupling will be investigated.

We are grateful to P. Ernstr\"om for useful discussions.
The work of L.B. was supported by the Swedish Natural Science Research
Council
(NFR) and EEC-SCIENCE contract no. SC1*-CT91-0650.
\newpage
{\Large Appendix}
\vskip .5cm
In this appendix, we prove the following theorem:

{\bf Theorem }
If $\chi(y)$ is a regular solution to (\ref{semiy}) with $\chi(0)=0$
then $0<|\chi(y)|<\pi$,
for all finite $y$, and we have $0<|\chi(\infty)|\leq\pi$.

For the proof we need some lemmas:

{\bf Lemma 1}
If $\chi(y)$ is a regular solution to (\ref{semiy}) with $\chi(0)=0$
then $0<|\chi(y)|<\pi$ for $0<y\leq1$.

The proof is similar to one used in \cite{stat} concerning
 the boundary conditions for the static ansatz $\chi(r,t)=f(r)$:

We make the variable substitution $x=1/y$ in (\ref{semiy})
and get,
\beq\label{semix}
(x^2-1)\chi_{xx}=\sin 2\chi(x),\; \chi(\infty)=0.
\eeq
We multiply this with $\chi_x$ and integrate from $x$ to $\infty$,
\beq\label{partint}
\left[(x^2-1)\chi_{x}^2\right]_x^\infty-2\int_x^\infty dx\:x\chi_x^2
=-\left[\cos 2\chi(x)\right]_x^\infty.
\eeq
Since $\chi_x(x)=-\frac{1}{x^2}(\chi_y(0)+{\cal O}(1/x))$ for large $x$,
(\ref{partint}) reduces to
\beq
-(x^2-1)\chi_{x}^2(x) - 2\int_x^\infty dx\:x\chi_x^2
=\cos 2\chi(x)-1.
\eeq
The left-hand side is always negative for $1\leq x<\infty$
so we must have
$0<|\chi(x)|<\pi$ for $1\leq x<\infty$.

{\bf Corollary}
If $\chi(y)$ is a regular solution to (\ref{semiy}) with $\chi(0)=0$
and $\chi_y(0)>0$ then $\chi(1)=\pi/2$, if $\chi_y(0)<0$ then
$\chi(1)=-\pi/2$.

This follows immediately from lemma 1 and the regularity condition
$\chi(1)=n \pi/2$, integer n.

{\bf Lemma 2}
If $\chi(y)$ is a solution to (\ref{semiy}) with $\chi(1)=\pi/2$ and
$\chi_y(1)>1$ then there exists a $0<y_0<1$ such that $\chi(y_0)=0$,
if $\chi_y(1)<-1$
 then there exists a $0<y_0<1$ such that $\chi(y_0)=\pi$.

A brief outline of the proof:

We make the substitution $y=\tan \theta$ which gives the equation
\beq
\cos 2\theta(\sin^2\theta \chi_{\theta\theta} +\sin 2\theta \chi_\theta)
=\sin 2 \chi(\theta).
\eeq
After differentiation of this equation one can get some inequalities on
the third derivative of $\chi$, these can then be used in order to show
that if $\chi_\theta(\pi/4)>2$ then
$\chi_\theta(\theta)>\chi_\theta(\pi/4)$, $0\leq\theta<\pi/4$.
( Note that $\chi_\theta(\pi/4=2\chi_y(1)$.)
This leads to the existence of a $0<\theta_0<\pi/4$ such that
$\chi(\theta_0)=0$.
A similar argument shows that if $\chi_\theta(\pi/4)<-2$ then
there exists a $0<\theta_0<\pi/4$ such that
$\chi(\theta_0)=\pi$.

{\bf Lemma 3}
If $\chi(y)$ is a solution to (\ref{semiy}) with $\chi(1)=\pi/2$
and $|\chi_y(1)|<1$ then $0<\chi(y)<\pi$, $y\geq 1$.

An outline of the proof:

We denote the known solutions with $\chi(1)=\pi/2$ by
$\chi^a(x)=\pi/2 \pm (2\: \arctan x-\pi/2)$.
Using the equation we get if we
multiply (\ref{semix}) with $\chi_x$ and integrate
 from $x$ to $1$, we can show that if $|\chi_x(1)|<1$ then
$|\chi_x(x)|<|\chi^a_x(x)|$ for $0\leq x\leq 1$.
Since $0\leq\chi^a(x)\leq\pi$
we thus have $0<\chi(x)<\pi$, $0\leq x\leq1$.
\emline

Proof of the theorem :

If $\chi(y)$ is a regular solution to (\ref{semiy}) with $\chi(0)=0$
and $\chi_y(0)>0$ then the corollary  tells us that $\chi(1)=\pi/2$.
 From lemma 1 together with lemma 2 it follows that
we can not have $|\chi_y(1)|>1$.
Together with the existence of the solution $ \chi(y)=2\:\arctan y$
(which has $\chi_y(1)=1$ and $\chi(\infty)=\pi$) we thus conclude that
 $-1<\chi_y(1)\leq1$.
{}From lemma 3 it now follows that $0<\chi(y)<\pi$, $y\geq1$,
this together with Lemma 1 thus tells us that $0<\chi(y)<\pi$, for all finite
$y$.
A similar reasoning gives $-\pi<\chi(y)<0$ if $\chi_y(0)<0$,
 which completes the proof.
\emline

\newpage
{\Large Figure Caption}
\vskip .5cm
\begin{itemize}
\item[Fig. 1]
 The previously known self-similar solution to the
non-linear sigma
model (solid line) and the first four of the new class of solutions found
in this paper. Here $\chi (y)$ is the radial function in the spherically
symmetric ansatz, and $y=r/t$is the self-similarity variable. The values
of the derivative at the origin are for each solution,
 $ \chi_y^0 =  2$,
 $ \chi_y^1=  21.757$,
 $ \chi_y^2=  234.50$,
 $ \chi_y^3=  2521.3$ and
 $ \chi_y^4=  27102$.
\end{itemize}
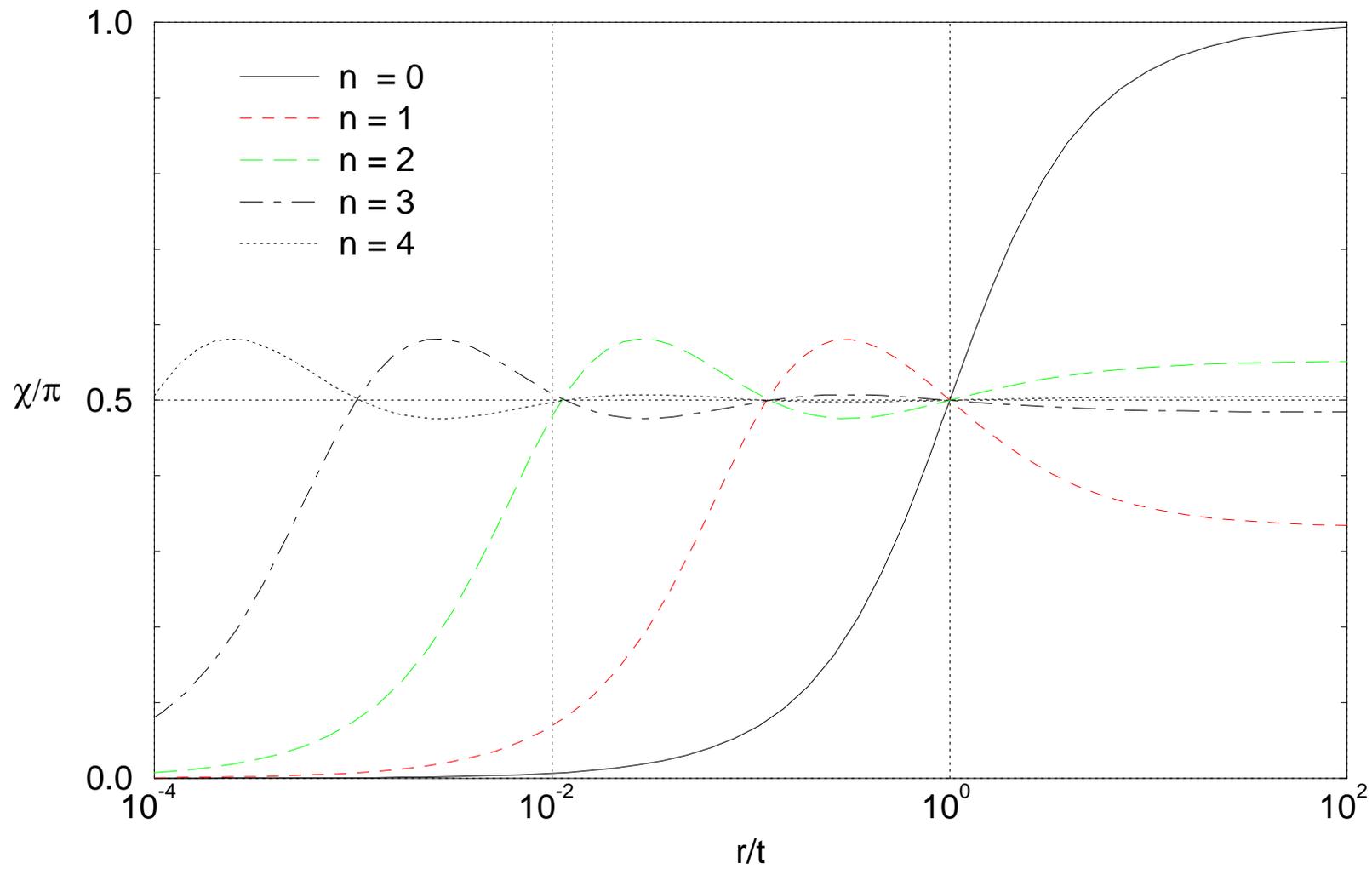
\end{document}